\documentclass[a4paper]{jpconf}
\usepackage{graphicx}
\usepackage{listings,multicol}
\usepackage{parcolumns}
\usepackage{xcolor}

\definecolor{dkgreen}{rgb}{0,0.6,0}
\definecolor{gray}{rgb}{0.5,0.5,0.5}
\definecolor{mauve}{rgb}{0.58,0,0.82}
\lstdefinestyle{myScalastyle}{
  frame=tb,
  float=*,
  language=scala,
  aboveskip=3mm,
  belowskip=3mm,
  showstringspaces=false,
  columns=flexible,
  basicstyle={\small\ttfamily},
  numbers=none,
  numberstyle=\tiny\color{gray},
  keywordstyle=\color{blue},
  commentstyle=\color{dkgreen},
  stringstyle=\color{mauve},
  frame=single,
  breaklines=true,
  breakatwhitespace=true,
  tabsize=3,
}

\begin{document}
\title[ACAT 2017: CMS Analysis and Data Reduction with Apache Spark]{CMS Analysis and Data Reduction with Apache Spark}

\author{Oliver Gutsche$^2$, Luca Canali$^1$, Illia Cremer$^4$, Matteo Cremonesi$^2$, Peter Elmer$^5$, Ian Fisk$^3$, Maria Girone$^1$, Bo Jayatilaka$^2$, Jim Kowalkowski$^2$, Viktor Khristenko$^1$, Evangelos Motesnitsalis$^1$, Jim Pivarski$^5$, Saba Sehrish$^2$, Kacper Surdy$^1$, Alexey Svyatkovskiy$^5$}

\address{$^1$European Organization for Nuclear Research CERN, Geneva, Switzerland}
\address{$^2$Fermi National Accelerator Laboratory, Batavia, IL, USA}
\address{$^3$Flatiron Institute of the Sions Foundation, New York, NY, USA}
\address{$^4$Intel Corp.}
\address{$^5$Princeton University, Princeton, NJ, USA}

\ead{gutsche@fnal.gov}

\begin{abstract}
Experimental Particle Physics has been at the forefront of analyzing the world's largest datasets for decades. The HEP community was among the first to develop suitable software and computing tools for this task. In recent times, new toolkits and systems for distributed data processing, collectively called ''œBig Data'' technologies have emerged from industry and open source projects to support the analysis of Petabyte and Exabyte datasets in industry. While the principles of data analysis in HEP have not changed (filtering and transforming experiment-specific data formats), these new technologies use different approaches and tools, promising a fresh look at analysis of very large datasets that could potentially reduce the time-to-physics with increased interactivity. Moreover these new tools are typically actively developed by large communities, often profiting of industry resources, and under open source licensing. These factors result in a boost for adoption and maturity of the tools and for the communities supporting them, at the same time helping in reducing the cost of ownership for the end-users. In this talk, we are presenting studies of using Apache Spark for end user data analysis. We are studying the HEP analysis workflow separated into two thrusts: the reduction of centrally produced experiment datasets and the end-analysis up to the publication plot. Studying the first thrust, CMS is working together with CERN openlab and Intel on the CMS Big Data Reduction Facility. The goal is to reduce 1 PB of official CMS data to 1 TB of ntuple output for analysis. We are presenting the progress of this 2-year project with first results of scaling up Spark-based HEP analysis. Studying the second thrust, we are presenting studies on using Apache Spark for a CMS Dark Matter physics search, investigating Spark's feasibility, usability and performance compared to the traditional ROOT-based analysis.
\end{abstract}


\section{Introduction}

Experimental Particle physics has been at the forefront of analyzing the world's largest datasets for decades. These datasets are produced by sophisticated detector systems that observe particle interactions. In this paper, we discuss the analysis of data from the field of High Energy Physics (HEP), where known particles are made to collide at the highest energies possible. The most basic concept of how data in HEP is organized is an event: all detector signals associated with a single beam crossing and subsequent high-energy collision. Events are the atomic unit of HEP data and may be processed separately, which is why the computational problems of particle physics can be easily parallelized. The underlying data organization holds for all sub-fields of experimental particle physics.

Events must be reconstructed to convert detector signals into measurements of particles produced in collisions. This is usually done centrally for each detector. The reconstructed events are then input to the final analysis done by individual researchers or groups of researchers exploring a multitude of physics questions. This process is very idiosyncratic, as individual researchers or groups are searching for different physics phenomena in the same data. This results in a challenging computational problem, where as many as thousands of physicists analyze the same datasets to extract different physics results.

The analysis process uses properties of the event such as energy or momentum of particles produced in the collisions. It is based on comparing the distribution of properties measured in many events with theoretical predictions of the same property, either calculated empirically or simulated using Monte Carlo techniques. Particle physics is a statistical science. Statistically significant numbers of recorded and simulated events are required to make claims.

Analysis is an iterative process. With ever increasing data volumes in particle physics, this process gets harder to perform interactively, simply due to the time required to read and transform the data. In the future, exponentially growing datasets will make this problem acute. New techniques will be required if we are to continue exploring the nature of matter and the universe.

Recently, new toolkits and systems have emerged outside of the HEP community to analyze Petabyte and Exabyte datasets in industry, collectively called 'Big Data'. These new technologies use different approaches and promise a fresh look at analysis of extremely large datasets. Our goal in investigating these new technologies is to improve the efficiency and turn-around times of HEP analysis. At the same time, using industry tools would educate our graduate students and postdoctoral researchers in these techniques and improve their chances on the job market outside academia. It would also make the HEP community part of an even larger community of data analysts.

In this paper, we present our experience applying one of these technologies, Apache Spark~\cite{Zaharia:2010:SCC:1863103.1863113}, to the HEP analysis problem. We incorporate lessons learned from our previous investigations~\cite{DBLP:journals/corr/GutscheCEJKPSSS17} and present new approaches and tools developed to enable HEP analysis in Apache Spark.

\section{The Traditional Analysis Workflow}
\label{sec:root_workflow}

The particle physics community was the first to develop suitable software and computing tools for the HEP analysis task. The requirement from the point of the analyst is to minimize the 'time-to-insight', to get from data to physics results as quickly as possible. Analysis is an iterative process, repeated in response to discoveries and mistakes. Interactivity is very important to facilitate an efficient conversation with the data. Asking scientific questions and answering them using data needs to be as user-friendly and direct as possible to be successful.

Many different physics topics can be under investigation concurrently, each looking at a different subset of the data. For all individual analyses, the same programmatic analysis steps are used to minimize 'time-to-insight': (a) Skimming (dropping events in a disk-to-disk copy), (b) Slimming (dropping variables in a disk-to-disk copy), (c) Filtering (selectively reading events into memory), (d) Pruning (selectively reading variables into memory).

The traditional user analysis workflow for CMS data applies two C++ frameworks to the centrally produced dataset: CMSSW~\cite{cmssw}, specially designed for processing CMS data, and ROOT~\cite{root}, which is a general, experiment-independent C++ toolkit. The ROOT framework provides statistical tools and a serialization format to persist reconstructed and transformed objects in files and is the underlying basis of CMSSW.

Although the CMSSW C++ framework can be very efficient, its data organization based on serialization of C++ classes in files can be difficult for end-users to understand. Moreover, it operates at a low level of abstraction, and even if a user knows how to process data efficiently, the difficulty in setting up a manual procedure may be outweighed by the time required to do so.

Most data analysts or analysis groups start by translating the class structure of the data into a "flat ntuple," in which events are rows of a table with primitive numbers or arrays of numbers as columns. These ntuples are written to files using the ROOT framework as well. They may then be analyzed without the CMSSW framework, with minimal dependencies.

This 'ntupling' step requires significant computing resources as it accesses the centrally produced samples. Mostly grid resources are used for this step to exploit parallelization. Workflow management systems take care of orchestrating the whole analysis workflow. Still significant burden of bookkeeping and failure re-submission is put on the individual analyst or analysis groups. 

Often, the ntuples are still too big for interactive analysis. Most analysis groups therefore introduce additional steps in which the ntuples themselves are skimmed and slimmed.

In the last step of the analysis, quantities from the final ntuple are aggregated and plotted as histograms. 

The time scale of a complete traditional analysis workflow
can range from days to weeks, depending on how many events are needed for the analysis. The first step is repeated about four times a year and the produced ntuples can be used by more than one analysis. The skimming and slimming step is executed between once and four times a month. The actual analysis step, producing plots, is repeated many times a day, since it represents the iterative and interactive analysis.


To give an example of data volumes of the different steps in a traditional analysis, we use a typical analysis of CMS~\cite{cms}, one of the 4 experiments at the Large Hadron Collider (LHC) at CERN in Geneva, Switzerland~\cite{1748-0221-3-08-S08001}. For an analysis of the 2015 dataset, the analysis ntuples have a size of about 2~Terabytes from a communal dataset of 200~Terabytes. These are then skimmed and slimmed down to several Gigabytes that are used for exploration and producing plots.

\section{The Spark Analysis Workflow}
\label{sec:spark_workflow}

In our previous usability study~\cite{DBLP:journals/corr/GutscheCEJKPSSS17} of Apache Spark, we implemented the traditional analysis workflow by converting data into the AVRO~\cite{avro} format and uploaded it to the HDFS~\cite{hdfs} file system of our development cluster. The analysis code itself was re-implemented in the scala~\cite{scala} language from it's original version written in C++. This resulted in a much simpler structure of the analysis itself due to the functional programming approach used in scala. It also resulted in much better portability to other facilities and technologies than in the traditional case.

While using Apache Spark to analyze data, analysts noticed that especially the tools to collect data in histograms were not suited for the new analysis paradigm. HEP physicists are used to control the event loop themselves and iterate through events programatically. Apache Spark manages the concurrency of the analysis and does not expose the event loop to the analysts. 

We solved this problem by developing a new toolkit especially designed for the map-reduce environment of Apache Spark and similar technologies. The {\bf Histogrammar}~\cite{jim_pivarski_2016_61418} package provides a functional interface to fill histograms in Apache Spark by passing lambda functions and use them in the same way as transformations are used in Apache Spark. The actual plot of a histogram is produced afterwards using a wide variety of plotting tools. 

The biggest impediment to use the new technology as identified by the analysts was the need to convert the data in a new format that Apache Spark can use efficiently. This step was considered time consuming and would not provide benefits over the traditional analysis workflow. To enable Apache Spark to understand the data structures of HEP experiments stored in ROOT files more directly, we developed a package called {\bf spark-root}~\cite{spark-root}. It is based on a Java-only implementation of the ROOT I/O libraries and connects ROOT to Apache Spark to be able to read ROOT object collections and automatically infer their class schema. The data is then available from within Apache Spark as DataFrames/Datasets/RDDs. This enables Apache Spark to read ROOT files directly from HDFS and also from the EOS~\cite{eos} storage system through a new Hadoop-xrootd connector~\cite{hadoop-xrootd-connector}. EOS is a storage system developed and deployed at CERN and other LHC grid sites. The connector was developed and tested with EOS. It is planned to expand its functionality gradually and support other storage system that allow access through the xrootd protocol~\cite{xrootd}.

In the following, we present performance tests of the full analysis workflow producing plots of histograms by reading ROOT files from the EOS storage system, which were conducted on CERN facilities.

A stand-alone test was performed using a dedicated test cluster of 3 nodes (1 name node, 2 data nodes) each with an Intel(R) Xeon(R) CPU E5-2650 @ 2.00GHz with 32 cores, 128 GB RAM and 10 Gb/s network connection. Files in different formats (text, parquet~\cite{parquet}, ROOT) were read either from the public or CERNBOX~\cite{cernbox} EOS instances at CERN, where the public EOS instance shows decreased performance characteristics because of its physical size and access pattern optimization compared to the CERNBOX instance. The results are shown in Table~\ref{tab:stand-alone}.

\begin{table}[htb!]
	\caption{Stand-alone test of reading files from the EOS storage system using a dedicated test cluster. Shown are copy and Spark access of different file formats. The access test of ROOT files from Spark used 2 executors and 4 cores.\\}
	\label{tab:stand-alone}
	\centering
	\begin{tabular}{l||l|l}
		\textbf{} & \textbf{EOS to HDFS copy} & \textbf{EOS from Spark}\\ \hline \hline
		             Text (200 GB from CERNBOX)    &              1 Gbit/s                        &              300 Mbit/s\\ \hline
		             PARQUET (200 GB from CERNBOX) &              800 Mbit/s                      &              6-9 Gbit/s\\ \hline
		             ROOT (200 GB from public EOS) &              400 Mbit/s &              2.3 Gbit/s\\
	\end{tabular}
\end{table}

The hadoop-xrootd connector and spark-root performed reasonably well in this stand-alone test. The copy test was limited by the performance of the public EOS instance performance and the characteristics of the HDFS setup of the test cluster, which copied a file first to the name node and then distributed it across the cluster. The Apache Spark test was limited by the uneven file size distribution of the input ROOT files and the fact that reading text files from Apache Spark is much less optimized then for example reading parquet files. The stand-alone tests didn't show any significant problems, although effects from different parts of the infrastructure (Apache Spark, local disk, memory, number of files, ...) are not easy to disentangle. Further tuning and optimization will be needed in the future.

We performed larger scale tests reading 0.5 TB of input ROOT files using the CERN analytix cluster of currently 36 nodes. A single node/single executor test was performed to compare reading ROOT files from HDFS and EOS. Reading 0.5 TB ROOT files from HDFS finished in about 4 hours where reading the same files from EOS only finished in about 9.5 hours. We concluded that the hadoop-xrootd connector is currently slower by a factor 2-3 than HDFS. It is expected that further tuning and optimization will be able to close the performance gap.

To test the parallel scaling by using more executors and cores of the analytix cluster, we distributed equal number of ROOT files over 160 tasks (40 executors with 4 cores per executor). This is sub-optimal because of the uneven distribution of file sizes, but sufficient for an initial large scale test. Table~\ref{tab:cluster} shows the results reading ROOT files from HDFS and EOS.

\begin{table}[htb!]
	\caption{Cluster test of reading ROOT files from HDFS and EOS. 160 tasks were used (40 executors with 4 cores per executor) to analyze 0.5 TB of ROOT files in Apache Spark. The total read time is calculated by subtracting the total CPU time from the total executor time.\\}
	\label{tab:cluster}
	\centering
	\begin{tabular}{l||l|l}
		\textbf{} & \textbf{HDFS from Spark} & \textbf{EOS from Spark}\\ \hline \hline
        Total executor time & 5.8 h& 11.7 h\\ \hline
        Total CPU time &2.9 h& 2.9 h\\ \hline
        Total Read time & 2.9 h&8.8 h\\ \hline
        Run time & 5 min&19 min\\
	\end{tabular}
\end{table}

\begin{figure}
\begin{center}
\includegraphics[width=\textwidth]{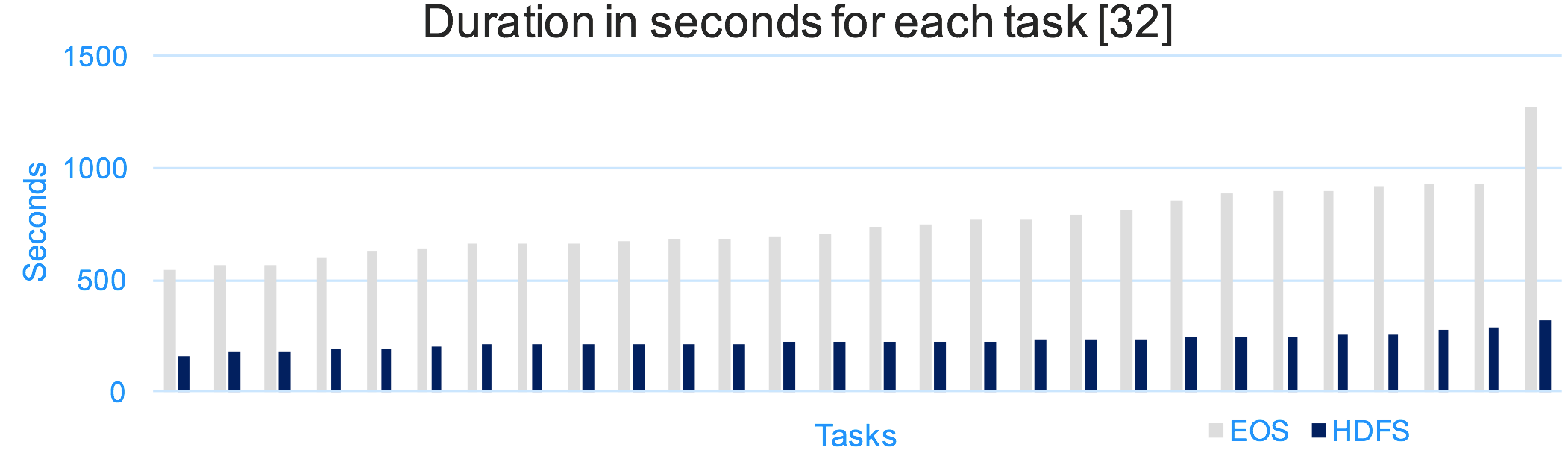}
\caption{\label{fig:executors}The parallel scalability test showed a larger difference between the HDFS and EOS read performance than the single executor/task test because of the tail of slow executors being longer for the xrootd-hadoop connector reading from EOS.}
\end{center}
\end{figure}

As expected we see a significant speed-up compared to the single executor/core case. The difference between HDFS and EOS read performance is larger than in the single case though. This is due to the uneven distribution of file sizes per task which results in the tail of slow executors being longer for the xrootd-hadoop connector (see for example Fig. \ref{fig:executors} in a different executor/core configuration with 32 tasks).

The average reading performance from HDFS is about 70 MB/s per task and from EOS about 20 MB/s per task. This confirms the conclusion that reading from EOS is currently about 2-3 times slower than from HDFS. The plan for the performance tuning is to first look at the impact of the network and optimize the task fragmentation.

\section{Conclusions and Next Steps}

We presented studies of executing the traditional HEP analysis workflow on Apache Spark. The two largest caveats from the initial studies were solved by developing the histogrammar package to fill histograms in map-reduce style technologies and to read ROOT files directly from Apache Spark, even from the CERN EOS storage system. This removed the need to convert the input data into a format that Apache Spark understands natively. 

The performance of the complete analysis workflow reading ROOT files is about 2-3 times slower on EOS than on HDFS, but further tuning and optimizations are expected to close the gap.

The scaling behavior results are promising to investigate larger and larger input volumes and reach the goal to perform interactive analysis on very large datasets. We hope the developed technologies and packages will enable the CERN openlab/Intel project~\cite{openlab} to develop a integrated analysis facility based on Apache Spark to reduce 1 PB of data in 5 hours to 1 TB for further analysis.

\section{Acknowledgments}

We would like to thank the CMS collaboration and the LHC to provide the data for the use case and the ROOT based workflow. We would also like to thank CERN openlab 
for enabling wide ranging collaboration on this research topic with international laboratories, universities and industry partners and being the spark for innovation. And we would like to thank Intel, our industry partner, in being interested in our challenges and helping us solving them.
This manuscript has been co-authored by Fermi Research Alliance, LLC under Contract No. DE-AC02-07CH11359 with the U.S. Department of Energy, Office of Science, Office of High Energy Physics, and by the National Science Foundation under grant ACI-1450377 and Cooperative Agreement PHY-1120138. The United States Government retains and the publisher, by accepting the article for publication, acknowledges that the United States Government retains a non-exclusive, paid-up, irrevocable, world-wide license to publish or reproduce the published form of this manuscript, or allow others to do so, for United States Government purposes.

\section*{References}
\bibliography{main}

\providecommand{\newblock}{}
\begin{thebibliography}{10}
\expandafter\ifx\csname url\endcsname\relax
  \def\url#1{{\tt #1}}\fi
\expandafter\ifx\csname urlprefix\endcsname\relax\def\urlprefix{URL }\fi
\providecommand{\eprint}[2][]{\url{#2}}

\bibitem{Zaharia:2010:SCC:1863103.1863113}
Zaharia M, Chowdhury M, Franklin M~J, Shenker S and Stoica I 2010 {\em
  Proceedings of the 2Nd USENIX Conference on Hot Topics in Cloud Computing\/}
  HotCloud'10 (Berkeley, CA, USA: USENIX Association) pp 10--10
  \urlprefix\url{http://dl.acm.org/citation.cfm?id=1863103.1863113}

\bibitem{DBLP:journals/corr/GutscheCEJKPSSS17}
Gutsche O, Cremonesi M, Elmer P, Jayatilaka B, Kowalkowski J, Pivarski J,
  Sehrish S, Surez C~M, Svyatkovskiy A and Tran N 2017 {\em CoRR\/} {\bf
  abs/1703.04171} \urlprefix\url{http://arxiv.org/abs/1703.04171}

\bibitem{cmssw}
Elmer P, Hegner B and Sexton-Kennedy L 2010 {\em J. Phys. Conf. Ser.\/} {\bf
  219} 032022

\bibitem{root}
Brun R and Rademakers F 1997 {\em Nuclear Instruments and Methods in Physics
  Research Section A\/} {\bf 389} 81 -- 86 ISSN 0168-9002 new Computing
  Techniques in Physics Research V
  \urlprefix\url{http://www.sciencedirect.com/science/article/pii/S016890029700048X}

\bibitem{cms}
Chatrchyan S~e~a (CMS Collaboration) 2008 {\em JINST\/} {\bf 3} S08004

\bibitem{1748-0221-3-08-S08001}
Evans L and Bryant P 2008 {\em Journal of Instrumentation\/} {\bf 3} S08001
  \urlprefix\url{http://stacks.iop.org/1748-0221/3/i=08/a=S08001}

\bibitem{avro}
Apache avro \urlprefix\url{http://avro.apache.org}

\bibitem{hdfs}
Shvachko K, Kuang H, Radia S and Chansler R 2010 {\em Proceedings of the 2010
  IEEE 26th Symposium on Mass Storage Systems and Technologies (MSST)\/} MSST
  '10 (Washington, DC, USA: IEEE Computer Society) pp 1--10 ISBN
  978-1-4244-7152-2 \urlprefix\url{http://dx.doi.org/10.1109/MSST.2010.5496972}

\bibitem{scala}
Scala \urlprefix\url{http://www.scala-lang.org}

\bibitem{jim_pivarski_2016_61418}
Pivarski J, Svyatkovskiy A, Schenck F and Engels B 2016 histogrammar-python:
  1.0.0 \urlprefix\url{https://doi.org/10.5281/zenodo.61418}

\bibitem{spark-root}
Khristenko V and Pivarski J 2017 diana-hep/spark-root: v0.1.14\_pre1 release
  \urlprefix\url{https://doi.org/10.5281/zenodo.1019880}

\bibitem{eos}
Eos: Large disk storage at cern \urlprefix\url{https://eos.web.cern.ch}

\bibitem{hadoop-xrootd-connector}
Motesnitsalis V hadoop-xrootd-connector
  \urlprefix\url{https://gitlab.cern.ch/awg/hadoop-xrootd-connector}

\bibitem{xrootd}
Dorigo A, Elmer P, Furano F and Hanushevsky A 2005 Xrootd- a highly scalable
  architecture for data access WSEAS Transactions on Computers

\bibitem{parquet}
Apache parquet \urlprefix\url{https://parquet.apache.org/}

\bibitem{cernbox}
Mascetti L, Labrador H~G, Lamanna M, Mościcki J and Peters A 2015 {\em Journal
  of Physics: Conference Series\/} {\bf 664} 062037
  \urlprefix\url{http://stacks.iop.org/1742-6596/664/i=6/a=062037}

\bibitem{openlab}
Cern openlab/intel cms big data project
  \urlprefix\url{https://cms-big-data.github.io}

\end{thebibliography}

\end{document}